\documentclass[usegraphicx,usenatbib]{mn2e}

\newcommand{\mh}{M_{\bullet}}
\newcommand{\milos}{Milosavljevi\'c\ }
\newcommand{\mdef}{M_{\rm def}}

\def\gap{\;\rlap{\lower 2.5pt
 \hbox{$\sim$}}\raise 1.5pt\hbox{$>$}\;}	
\def\lap{\;\rlap{\lower 2.5pt	
   \hbox{$\sim$}}\raise 1.5pt\hbox{$<$}\;}
	
\begin{document}
\title[Galaxy Cores]{Galaxy Cores as Relics of Black Hole Mergers}
\author[\milos {\it et al.}]
{Milo\v s Milosavljevi\' c$^1$, David Merritt$^1$, Armin Rest$^2$, and Frank C. van den Bosch$^3$ \\
$^1$Department of Physics and Astronomy, Rutgers University, Piscataway, NJ, 08854, USA \\
$^2$ Department of Astronomy, University of Washington, Seattle, WA, 98195, USA \\
$^3$ Max-Planck Institut f\"ur Astrophysik, Karl-Schwarzschild Strasse 1, Postfach 1317, 85741 Garching, Germany}

\maketitle

\begin{abstract}
We investigate the hypothesis that the cores of elliptical galaxies and 
bulges are created from the binding energy liberated by the coalescence of 
supermassive binary black holes during galaxy mergers.
Assuming that the central density profiles of galaxies were
initially steep power laws, $\rho\sim r^{-2}$, we define the
``mass deficit'' as the mass in stars that had to be removed from 
the nucleus in order to produce the observed core.
We use nonparametric deprojection to compute the mass deficit in a 
sample of 35 early-type galaxies with high-resolution imaging data.
We find that the mass deficit correlates well with the mass of the nuclear
black hole, consistent with the predictions of merger models.
We argue that cores in halos of non-interacting dark matter
particles should be comparable in size to those observed in the stars.
\end{abstract}

\section{Introduction}
\citet{fer94} and \citet{lau95} 
divide elliptical galaxies
into two classes based on their nuclear properties, which Lauer {\it et al.}
call ``core'' and ``power-law'' galaxies.
Core galaxies exhibit a definite break in the surface brightness profile
at some radius $R_b$; inward of this break, the logarithmic slope
gently decreases in a manner that mimics a constant-density core.
Power-law galaxies show essentially a single power-law profile
throughout their inner regions, $\Sigma(R)\sim R^{-\Gamma}$, 
$\Gamma\approx -0.8\pm 0.2$.
The brightest galaxies, $M_V\lap -21$, are exclusively core galaxies
while faint galaxies, $M_V\gap -16$ always exhibit power laws;
galaxies of intermediate luminosity can exhibit either type of profile
\citep{geb96}.
While the two categories were initially seen as distinct,
nonparametric deprojection revealed that even the ``core'' galaxies
exhibit power laws in their central space densities,
$\rho\sim r^{-\gamma}$, with $\gamma\lap 1$ \citep{mef95}.
Power-law galaxies have $1\lap\gamma\lap 2.5$ \citep{geb96}.
Furthermore the distribution of de-projected slopes is essentially 
continuous as a function of galaxy luminosity in the larger samples 
now available \citep{rav01, res01}.

Here we assume that the steep central density cusps of faint ellipticals
and bulges, $\rho\sim r^{-2}$, are characteristic of the earliest generation 
of galaxies, and ask: How do the low-density cores associated with bright 
galaxies form?
An appealing hypothesis links cores to nuclear black holes (BHs): 
in a galactic merger, the BHs will fall to the center of the merger remnant
and form a bound pair, releasing their binding energy to the surrounding
stars \citep{bbr80,ebi91}.
High-resolution $N$-body simulations verify that this process can
convert a steep power-law cusp, $\rho\sim r^{-2}$, into a shallow
power-law cusp, $\rho\sim r^{-1}$, within the radius of gravitational
influence of the BHs \citep{mme01}.
Successive mergers would presumably lower the density of the core still
more. 
In this model, power-law galaxies are those which have not experienced
a major merger since the era of peak BH growth, or which
have re-generated their cusps via star formation \citep{mme01}.

Preliminary tests of the cusp-disruption model were presented by 
\citet{fab97} and \citet{mme01}.
The former authors plotted core properties (break radius,
core luminosity) versus global properties
in a sample of 19 early-type galaxies and noted a rough proportionality.
In the paradigm investigated here, 
core properties should correlate more fundamentally with BH
mass, since the mass of stars ejected by a decaying binary BH is expected
to be of order the BHs' mass.
\citet{mme01} used the new empirical relation between
galaxy velocity dispersion and BH mass, the $\mh-\sigma$ relation
\citep{fem00,geb00}, to estimate BH masses in the Faber
{\it et al.} sample.  They found that a rough dynamical estimate of the 
``mass deficit'' -- the mass that would need to
be removed from an initially $r^{-2}$ density cusp in order to produce
the observed profile -- correlated well with $\mh$.  

In this paper we present the most careful test to date of the BH
merger hypothesis for the formation of galaxy cores.
We use nonparametric deprojection
to compute the mass deficit in a sample of galaxies with high-resolution 
imaging data from HST (\S2).  We find a strong correlation 
between this mass and the mass of the nuclear BH; typical
ejected masses are $\sim$ several $\mh$ (\S3).
We argue (\S4) that this result is consistent with the formation of
cores via hierarchical mergers of galaxies containing pre-existing BHs.
Cusps of non-interacting dark matter particles should behave in 
the same way as cusps of stars in response to heating by binary BHs,
and we argue that the damage done to stellar cusps by this mechanism
is a reasonable guide to the damage that would be done to dark matter cusps.

\section{Data and Method}
\label{sec_data}
Our data set is drawn from a sample of 67 surface brightness profiles of 
early-type galaxies observed with HST/WFPC2 by \citet{res01},
and three additional galaxies: NGC~4472 and 4473, observed with
WFPC1 by \citet{fer94}; and a WFPC2 F547M image of M87 
\citep{jor02}. 
The Rest {\it et al.} sample was selected from the set of all early-type
galaxies with radial velocities less than 3400 km s$^{-1}$,
absolute $V$-band magnitudes less than $-18.5$, and absolute galactic 
latitude exceeding 20 degrees.
From this sample we excluded 13 galaxies for which central velocity 
dispersions were not available in literature; as discussed below,
velocity dispersions were needed to compute BH masses in most of the galaxies. 
For specifics of the image-data reduction we refer the reader to the 
sources cited above.  Surface brightness profiles used in this study were 
major-axis profiles.  We applied a crude correction for the apparent 
ellipticity of the galaxies by multiplying the volume-integrated quantities, 
defined below, by $(1-\epsilon_b)$, where $\epsilon_b$ is the 
ellipticity of the isophote at the break radius.

The intrinsic luminosity profiles $\nu(r)$ were obtained by deprojecting 
the PSF-deconvolved surface brightness profiles $\Sigma(R)$ using the 
non-parametric MPL technique \citep{met94}.  
We opted for one-step deprojection via maximization of the penalized
likelihood functional
\begin{equation} 
\label{eq_mpl} {\mathcal L}_\lambda [\nu] =
\sum_i \frac{\left(\Sigma_i - P_i[\nu] \right)^2}{\left(\Sigma^{err}_i\right)^2}
-\lambda \int_0^\infty \left[\frac{d^2\log\nu}{(d\log r)^2}\right]^2 d\log r.
\end{equation} 
The first term compares the observed surface brightness data
$\Sigma_i\equiv\Sigma(R_i)$ to the projections $P_i[\nu]$ of the intrinsic luminosity estimate $\nu$ given by the operator
\begin{equation} 
P_i[\nu] = 2 \int_{R_i}^\infty
\frac{\nu(r) r dr}{\sqrt{r^2-{R_i}^2}}.
\end{equation} 
The second term in equation \ref{eq_mpl} is the penalty function which assigns 
zero penalty to any power-law $\nu(r)$;
hence our estimate of $\nu$ is unbiassed if $\nu$ is an unbroken power law
and should be minimally biassed if $\nu$ is approximately a power law.
The relative strength of the penalty term is regulated through the parameter 
$\lambda$ which was chosen by eye and equal to $0.01$ for all galaxies; 
integrated quantities like the mass deficit
defined below are only weakly dependent on $\lambda$.

Distances for 26 of the galaxies were drawn from the SBF survey \citep{ton01}.
For the remaining 34 galaxies we adopted distances computed assuming
a pure Hubble expansion with $H_0=80\textrm{ km s}^{-1}\textrm{ Mpc}^{-1}$ 
corrected for Virgo-centric infall \citep{res01}.
Luminosity densities were converted to mass densities
$\rho(r)=\Upsilon\nu(r)$ using the individual mass-to-light ratios 
$\Upsilon_V$ quoted in \citet{mag98} 
or their best-fit relation $\log(\Upsilon_V/\Upsilon_{\odot,V})=
-1.11\pm0.33 +(0.18\pm0.03)\log(L_V/L_{\odot,V})$ for galaxies not 
included in that study.

\begin{figure}
\includegraphics[width=8.0cm]{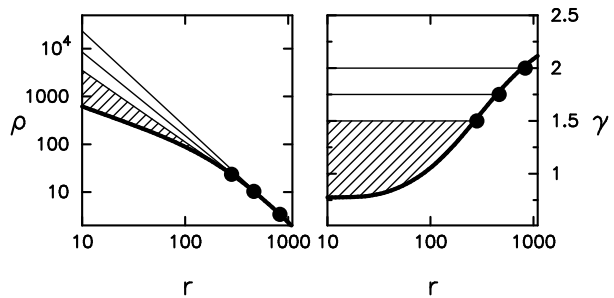}
\caption{Definition of the mass deficit, illustrated using the profile
of NGC~5903.
Left panel: density in $M_\odot\textrm{pc}^{-3}$ as a function of radius 
in parsecs (thick line); hypothetical original density cusps 
for fiducial slopes $\gamma_0=2, 1.75, 1.5$ (thin lines). 
Dots indicate the corresponding break radii.
Shaded region is the mass deficit for $\gamma_0=1.5$.  
Right panel: the negative logarithmic derivative
of the density profile, $\gamma\equiv-d\log\rho/d\log r$.
Break radii are defined as the radii where $\gamma=\gamma_0$.}
\label{fig_defin}
\end{figure}

We define $\gamma\equiv-d\log\rho(r)/d\log(r)$ as the local, negative
logarithmic slope of the deprojected density profile.  
Power-law galaxies are defined as those in which $\gamma\ge 2$ at
all radii; typically the profiles of such galaxies show no clear
feature that can be identified as a ``break radius'' and are unlikely
candidates for cusp destruction by binary BHs.  
In the remaining 35, ``core'' galaxies, the slope varies from $\gamma>2$
at large radii to $\gamma<2$ at small radii; the radius at which
the slope crosses $\gamma=2$ in the positive sense ($d\gamma/dr>0$) is
called here the ``break radius'' $r_b$ (Figure \ref{fig_defin}).  This definition has 
little in common with the more standard definition 
based on fitting of the {\it surface brightness} profile to an ad hoc
parametric function.
In four galaxies the slope crosses $\gamma=2$ in the positive sense at 
more than one radius and thus the definition of $r_b$ is ambiguous.  
In such cases, we select the crossing toward larger radius
from the largest dip of the slope below $\gamma=2$.  

We define the {\it mass deficit} as the difference in integrated mass 
between the deprojected density profile $\rho(r)$ and a $\gamma=\gamma_0=2$ 
profile extrapolated inward from the break radius:
\begin{equation}
\mdef\equiv 4\pi (1-\epsilon_b) \int_0^{r_b} 
\left[\rho(r_b) \left(\frac{r}{r_b}\right)^{-\gamma_0} -\rho(r)\right] r^2 dr.
\end{equation}
Our choice of an $r^{-2}$ density profile to characterize the
``undisrupted'' core is to a certain extent arbitrary;
adiabatic growth of BHs can produce cusps with $1.5\lap\gamma\lap 2.5$ 
depending on initial conditions, and the faintest ellipticals with
measured cusp slopes exhibit a similar range of $\gamma$'s (e.g.,
\citet{geb96}).
To test the sensitivity of our results to the assumed initial profile,
we repeated the analysis using fiducial slopes of $\gamma_0=1.75$ and
$1.5$.
Values of $\gamma_0>2$ were found to exclude all but a few galaxies.

BH masses for a few of the galaxies in our sample are
available from spatially-resolved kinematical studies
\citep{mef01}.
For all other galaxies we estimated $\mh$ via the $\mh-\sigma$ relation,
\begin{equation}
\label{eq_msigma}
\mh\approx 1.4\times10^8 M_\odot 
\left(\frac{\sigma_c}{200\textrm{ km s}^{-1}}\right)^{4.8\pm0.5}
\end{equation}
\citep{fem00}, where $\sigma_c$ is the central
velocity dispersion corrected to an aperture of $r_e/8$ with $r_e$
the effective radius.
We used the aperture corrections of \citet{jor95}
to compute $\sigma_c$ from published values of $\sigma$ in
\citet{dav87}, \citet{ton81} and \citet{din95}.

\section{Results}
\begin{figure}
\includegraphics[width=8.0cm]{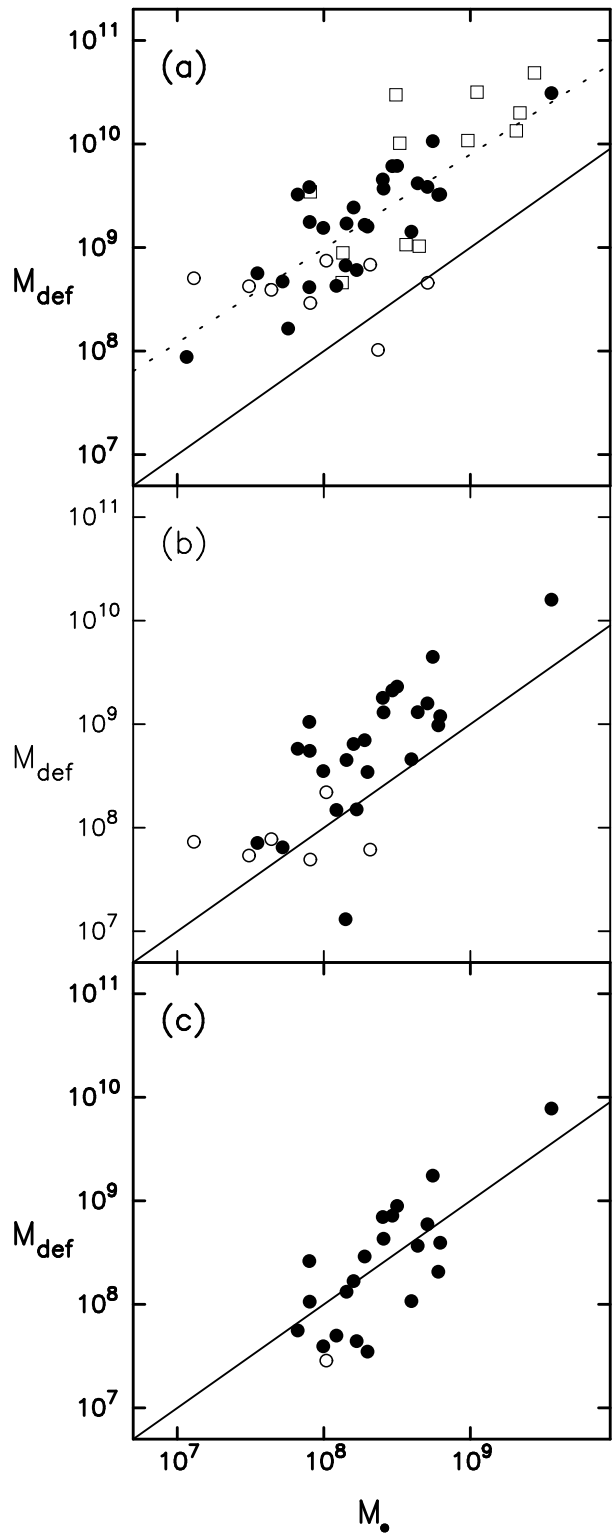}
\caption{Mass deficit vs. BH mass for three different values of $\gamma_0$, 
the assumed logarithmic slope of the density cusp before energy input
from BHs. 
(a) $\gamma_0=2$; (b) $\gamma_0=1.75$; (c) $\gamma_0=1.5$.  
Filled circles are elliptical galaxies and empty circles are lenticulars. 
Squares show dynamical estimates of $\mdef$ for the sample
of galaxies considered in \citet{mme01}.
Solid line is the $\mdef=\mh$ relation.
Dotted line in panel (a) is a linear regression 
fit to all galaxies (see Table \ref{tab_slopes}).}
\label{fig_data} 
\end{figure}

Results are given in Tables \ref{tab_galaxies} and \ref{tab_slopes} and Figure \ref{fig_data}; 
Table \ref{tab_galaxies} gives
mass deficits only for $\gamma_0=2$ while Figure \ref{fig_data} shows
$\mdef$ computed using all three values of the fiducial slope,
$\gamma_0$=(2, 1.75, 1.5).
In the first panel of Figure \ref{fig_data} 
we have also plotted ``dynamical'' estimates of
$\mdef$ for a sample of galaxies from \citet{geb96}, 
using equation (41) from \citet{mme01},
$M_{\rm dyn}\equiv2(2-\gamma)/(3-\gamma)\sigma^2R_b/G$;
$M_{\rm dyn}$ depends on the density profile only through $R_b$, the
break radius of the surface bightness profile, and $\gamma$.
When calculated for the deprojected galaxies in our sample,
$M_{\rm dyn}$ was consistent within the scatter with $\mdef$.

For $\gamma_0=2$,
the mass deficits are clustered about a linear relation defined by 
$\langle \log (\mdef/\mh) \rangle = 0.92$, $1.0$, and
$0.65$, respectively, for Es and S0s; Es only; and S0s only,
corresponding to $\mdef\sim (8.4,10,4.5)\mh$.
Decreasing $\gamma_0$ decreases $\mdef$ (cf. Fig. 1b) and $\mdef$
becomes negative/undefined in galaxies when the minimum pointwise 
slope $\gamma_{\rm min}$ approaches $\gamma_0$.
We do not cite values of $\langle\log(\mdef/\mh)\rangle$ for these low values 
of $\gamma_0$ since the mean depends strongly on which galaxies
are defined as having ``cores.''
However Figures \ref{fig_data}b,c shows that $\mdef$ remains 
of order $\mh$ or greater
for many of the galaxies even when $\gamma_0<2$.
We speculate that the lower mean value of $M_{\rm def}$ for the 
S0s may indicate a role for gaseous dissipation in the 
re-formation of cusps following mergers.  We emphasize, however, 
that the mass deficit of lenticulars alone appears to be completely 
uncorrelated with the BH mass.

Table \ref{tab_slopes} lists linear regression fits to the data 
that were carried out using the routine of \citet{akb96}.  
For the \citet{res01} ellipticals, 
as well as for the complete sample, the fitted power-law indices
of the $\mdef-\mh$ relation calculated with 
$\gamma_0=2.0$ and $1.75$ are statistically consistent with unity.  
To test the sensitivity of the fitting parameters to the assumed 
power-law index $\alpha$ of the $\mh-\sigma$ relation 
(equation \ref{eq_msigma}), 
we recomputed the fits with the \citet{geb00} value $\alpha=3.75\pm0.3$ 
in addition to the \citet{fem00} value $\alpha=4.85\pm0.5$ 
that is standard throughout the current paper.  The effect of changing to 
a shallower version of the $\mh-\sigma$ relation is 
a steepening of the $\mdef-\mh$ relation, 
from $d\log\mdef/d\log\mh\approx 0.91$ to $d\log\mdef/d\log\mh\approx 1.16$.

We note that in most galaxies the break radius defined via 
$\gamma(r_b)=\gamma_0=2$ is close to the radius at which the 
density profile exhibits a visual break, i.e., where the curvature is
greatest.  The break radius, however, is not a good predictor
of the mass deficit; in particular, some galaxies with large break radii, 
$r_b\gg r_\bullet$, with $r_\bullet=G\mh/\sigma^2$ the dynamical
radius of the BH, have mass deficits below the mean.
We speculate below that these large break radii may be produced
by mechanisms other than BH mergers.

Although estimation of errors in Figure \ref{fig_data} is difficult, 
we believe that the scatter is at least partly intrinsic.
Uncertainties in $\log\mh$ are due primarily to uncertainties in
$\sigma$ and are of order $\log 2$.  Uncertainties in $\log\mdef$ are
also roughly $\log 2$ based on variances between the redshift-based and SBF
distances.
A scatter greater than that due to measurement uncertainties would
be reasonable given the different merger histories of galaxies
with given $\mh$.  Similarly, if the progenitors of the galaxies 
exhibited a range of different $\gamma_0$, this could in itself explain 
the observed scatter in Figure \ref{fig_data}.

\section{Discussion}
The mass ejected by a decaying BH binary is
\begin{equation}
M_{\rm ej} \approx J M_{12} \ln\left({a_h\over a_{gr}}\right)
\end{equation}
where $M_{12}=M_1+M_2$ is the binary mass,
$a_h$ is the semi-major axis when the binary first becomes hard, and
$a_{gr}$ is the separation at which the rate of energy loss to gravitational
radiation equals the rate of energy loss to the stars \citep{qui96}.
$J$ is a dimensionless mass-ejection rate;
for equal-mass binaries, $J\approx 0.5$ \citep{mme01}.
\citet{qui96} claims that $J$ is nearly independent of $M_2/M_1$
even for extreme mass ratios, implying that a mass in stars
of order $M_1+M_2$ is ejected during {\it every} accretion event, 
even when $M_2$ is tiny compared with $M_1$.
This non-intuitive result is due to Quinlan's ejection criterion,
which for $M_2\ll M_1$ includes stars that would not have gained
sufficient energy to escape from the binary.
If instead we equate the change in energy of the binary
with the energy carried away by stars that are ejected
with $v\gap V_{bin}$,
we find $M_{ej}\approx M_2$.
This argument suggests a relation
\begin{equation}
M_{\rm ej} \approx M_2 \ln\left({a_h\over a_{gr}}\right)
\end{equation}
which is consistent with equation (5) when $M_1\approx M_2$.

Using equation (6),
and adopting Merritt's (2000) semi-analytic model
for decay of a binary in a power-law cusp, we find
\begin{equation}
{a_{gr}\over a_h}\approx A\left|\ln A\right|^{0.4}, \ \ \ 
A \approx 7.5\left({M_1\over M_2}\right)^{0.2}{\sigma\over c}
\end{equation}
and
\begin{equation}
M_{ej}\approx 4.6 M_2\left[1 + 0.043\ln\left({M_2\over M_1}\right)\right], \ \ \ M_1\le M_2.
\label{eq_mej}
\end{equation}
Thus $M_{ej}/M_2$ varies only negligibly with $M_1/M_2$.
Henceforth we adopt $M_{ej}\approx 5M_2$.

If a BH grows by sequential accretion of smaller BHs, this
result implies a mass deficit of order five times the
final BH mass.
However if the BH grows via a merger hierarchy of comparably-massive
BHs, we expect $M_{def}$ to be larger.
The idea here is that the damage done to cusps is {\it cumulative}:
a merger of two galaxies whose cusps had previously been destroyed
by binary BHs, will produce a shallower profile than a merger
between two galaxies with initially steep cusps, even if the final
BH mass is the same.
Galaxies with masses $M\gap 10^{11}M_{\odot}$,
including most of the galaxies plotted on Figure \ref{fig_data}, 
are believed
to have undergone at least one major merger since a redshift of
$1$ (e.g. Kauffmann, Charlot \& Balogh 2001).  Thus we predict $\mdef\gap 5\mh$, 
consistent with Figure \ref{fig_data} if $\gamma_0\gap1.5$.

Our interpretation of the mass deficit depends critically on the 
assumption that all of the change in $\gamma$ during a merger 
can be attributed to the BHs,
i.e., that cusp slopes remain unchanged the absence of BHs.
This is known to be the case in equal-mass mergers between galaxies with
power-law cusps \citep{bar99, mme01},
though in mergers with extreme mass ratios, features can appear in the
density profile that are not due to BHs \citep{mec01}.
We speculate that the break radii in some of the galaxies in our sample
may be due to this process, particularly those galaxies (e.g., NGC~3640, 4168)
where $r_b$ greatly exceeds the radius of gravitational influence of
the BH.
$N$-body simulations of cumulative mergers with unequal BH masses 
will be needed to assess this hypothesis.

Our model presents an interesting contrast to
that of \citet{vdm99}, who proposed that cores 
(in the sense of constant-density regions) were present
in all galaxies ab initio, and that power-law cusps were 
generated by the growth of the BHs -- roughly the opposite
of our model in which BHs destroy pre-existing cusps.  Van der Marel
assumed that core mass correlated initially with bulge 
luminosity as $M_{\rm core}\sim L^{1.5}$ and that 
$M_{\bullet}\propto L$; hence $M_{\rm core}\propto M_{\bullet}^{1.5}$,
consistent with the correlation in Figure \ref{fig_data} if we identify
$M_{\rm core}$ with $\mdef$. We believe that this agreement is coincidental.  
Van der Marel's model relates core mass to BH mass via an ad hoc
postulate, while the model discussed here contains a mechanism for 
core formation.
Van der Marel also ignored the effects of mergers.
Nevertheless, van der Marel's model shows that our interpretation
is not unique.

If stellar cusps are destroyed by binary BHs, the same should
be true of dark-matter cusps, like those predicted in
CDM theories of structure formation (e.g., \citet{nfw96, moo98, bul01}). 
Destruction of dark matter cusps could be very efficient if
supermassive BHs were present in dark matter halos
at large redshifts (e.g. \citet{fan01, hal01, mhn01})
due to the cumulative effect mentioned above;
furthermore, dark matter cusps would not be regenerated the way 
that stellar cusps might be via star formation.
If our model for the formation of stellar cores is correct,
we would predict tha cores of non-interacting CDM 
should be about as 
large as those observed in the stars, and perhaps larger.

We thank L. Ferrarese for her generous help with the data
reduction and interpretation.  The image of M87 used for 
calculating the mass deficit of that galaxy (Section \ref{sec_data})
was kindly provided
by P. C\^ot\'e and A. Jord\'an in advance of publication.
This work was supported by NSF grant AST 00-71099 and NASA grants
NAG5-6037 and NAG5-9046 to DM.

\bibliographystyle{mn2e.bst} 
\bibliography{cores.bbl}

\begin{table}
\begin{tabular}{lcccccc}
 Galaxy  & 
 $D$ & 
 $M_B$ &
 $r_b$ &
 $\gamma_{\rm min}$ & 
 $\log \mh$ &
 $\log \mdef$ \\
\hline
NGC~2549 & 12.6 & -18.5 & 0.22 & 1.59 & 7.91 & 8.46  \\
NGC~2634 & 33.4 & -19.9 & 0.27 & 1.73 & 7.90 & 8.62  \\
NGC~2986 & 26.8 & -20.5 & 0.41 & 0.99 & 8.79 & 9.51  \\
NGC~3193 & 34.0 & -20.9 & 0.23 & 0.84 & 8.15 & 9.23  \\
NGC~3348 & 38.5 & -21.1 & 0.49 & 0.75 & 8.47 & 9.79  \\
NGC~3414 & 25.2 & -20.1 & 0.14 & 1.61 & 8.71 & 8.66  \\
NGC~3613 & 29.1 & -20.7 & 0.52 & 0.80 & 8.20 & 9.39  \\
NGC~3640 & 27.0 & -21.0 & 0.68 & 0.94 & 7.82 & 9.51  \\
NGC~4121 & 28.3 & -17.9 & 0.12 & 1.76 & 7.06 & 7.94  \\
NGC~4128 & 32.4 & -19.8 & 0.18 & 1.64 & 8.32 & 8.83  \\
NGC~4168 & 30.9 & -20.4 & 1.25 & 1.01 & 7.90 & 9.58  \\
NGC~4291 & 26.2 & -19.8 & 0.17 & 0.49 & 8.28 & 9.22  \\
NGC~4365 & 20.4 & -21.1 & 0.57 & 0.71 & 8.50 & 9.79  \\
NGC~4472 & 16.3 & -21.7 & 0.63 & 0.90 & 8.74 & 10.03 \\
NGC~4473 & 15.7 & -19.9 & 0.24 & 1.22 & 7.90 & 9.25  \\
NGC~4478 & 18.1 & -19.1 & 0.30 & 1.42 & 7.55 & 8.75  \\
NGC~4486 & 16.1 & -21.5 & 1.12 & 0.74 & 9.55 & 10.49 \\
NGC~4503 & 17.6 & -19.2 & 0.25 & 1.56 & 7.11 & 8.70  \\
NGC~4564 & 15.0 & -18.9 & 0.11 & 1.69 & 7.76 & 8.22 \\
NGC~4589 & 22.0 & -20.0 & 0.18 & 1.05 & 8.22 & 8.78 \\
NGC~5077 & 34.0 & -20.4 & 0.54 & 1.15 & 8.78 & 9.51 \\
NGC~5198 & 34.1 & -20.0 & 0.11 & 0.92 & 8.09 & 8.63 \\
NGC~5308 & 26.6 & -19.7 & 0.10 & 1.84 & 8.37 & 8.01 \\
NGC~5370 & 41.3 & -19.0 & 0.32 & 1.56 & 7.49 & 8.63 \\
NGC~5557 & 42.5 & -21.2 & 0.44 & 0.87 & 8.64 & 9.62 \\
NGC~5576 & 25.5 & -20.3 & 0.24 & 1.37 & 8.00 & 9.19 \\
NGC~5796 & 36.5 & -20.7 & 0.18 & 1.18 & 8.60 & 9.15 \\
NGC~5812 & 26.9 & -20.3 & 0.25 & 1.71 & 8.15 & 8.83 \\
NGC~5813 & 32.2 & -21.1 & 0.43 & 0.31 & 8.40 & 9.66 \\
NGC~5831 & 27.2 & -19.9 & 0.22 & 1.42 & 7.72 & 8.67 \\
NGC~5898 & 29.1 & -20.4 & 0.33 & 1.34 & 8.30 & 9.20 \\
NGC~5903 & 33.9 & -20.9 & 0.78 & 0.77 & 8.41 & 9.57 \\
NGC~5982 & 39.3 & -20.9 & 0.45 & 0.49 & 8.71 & 9.58 \\
NGC~6278 & 37.1 & -19.7 & 0.10 & 1.53 & 7.64 & 8.59 \\
UGC~4551 & 23.6 & -18.7 & 0.24 & 1.36 & 8.02 & 8.87 \\
\hline
\end{tabular}
\caption{Galaxies with $\gamma_{\rm min}\leq 2$;
$D$ is distance in Mpc; $r_b$ is break radius in kpc; 
$\gamma_{\rm min}$ is the minimum logarithmic slope; 
$\mh$ and $\mdef$ are in solar masses.}
\label{tab_galaxies}
\end{table}

\begin{table}
\begin{tabular}{ccccc}
 $\gamma_0$  &
	     E  &
             S0  & 
	     $\alpha=4.8$  & 
	     $\alpha=3.75$ \\
\hline
$2.00$ & $0.93\pm0.10$ & $-0.10\pm0.14$ & $0.91\pm0.09$ & $1.16\pm0.12$ \\
       & $(10.5\pm1.8)$ & $(3.8\pm 0.86)$ & $(9.75\pm1.78)$ & $(10.7\pm1.7)$ \\
$1.75$ & $1.07\pm0.15$ & $-0.78\pm0.46$ & $0.86\pm0.24$ & $1.10\pm0.30$ \\
       & $(2.75\pm 0.78)$ & $(0.35\pm0.19)$ & $(2.17\pm0.50)$ & $(2.36\pm0.56)$ \\
$1.50$ & $1.56\pm0.37$ &  & $1.58\pm0.35$ & $2.02\pm 0.45$ \\
       & $(0.49\pm0.26)$ &  & $(0.46\pm0.23)$ & $(0.55\pm0.25)$  \\
\hline
\end{tabular}
\caption{Linear regression fits to the 
($\log\mh,\log\mdef$) data for three 
values of the fiducial logarithmic slope $\gamma_0$.  
Values in parentheses are 
$\mdef/\mh$ interpolated from 
the fit at $\mh=10^8M_\odot$.  Fourth and fifth columns are, respectively, 
fits of the entire data set (including the galaxies 
with dynamical estimates of $\mdef$) using the \citet{fem00}
and the \citet{geb00} values of 
the $\mh-\sigma$ relation exponent $\alpha$.}
\label{tab_slopes}
\end{table}

\end{document}